\documentstyle[aps,multicol,prb,epsfig]{revtex}
\begin{document}
\draft
\title{
Damped orbital excitations in the titanates
}
\author{K. Kikoin$^{a}$, O. Entin-Wohlman$^{b}$, V.
Fleurov$^{b}$, and A. Aharony$^{b}$}
\address{$^a$Department of
Physics, Ben Gurion University, Beer Sheva 84105, Israel}
\address{$^b$School of Physics and Astronomy, Raymond and Beverly
Sackler Faculty of Exact Sciences, \\ Tel Aviv University, Tel
Aviv 69978, Israel\\ }

\date{\today}
\maketitle
\begin{abstract}

A possible mechanism for the removal of the orbital degeneracy in
RTiO$_{3}$ (where R=La, Y, ...) is considered. The calculation is
based on the Kugel-Khomskii Hamiltonian for electrons residing in
the $t_{2g}$ orbitals of the Ti ions, and uses a self-consistent
perturbation expansion in the interaction between the orbital and
the spin degrees of freedom. The latter are assumed to be ordered
in a N\'eel state, brought about by delicate interactions that are
not included in the Kugel-Khomskii Hamiltonian. Within our model
calculations, each of the $t_{2g}$ bands is found to acquire a
finite, temperature-dependent dispersion, that lifts the orbital
degeneracy.
The orbital excitations are found to be heavily damped over a
rather wide band. Consequently, they do not participate as a
separate branch of excitations in the low-temperature
thermodynamics.
\end{abstract}


\begin{multicols}{2}

\section{Introduction}

The Mott insulator lanthanum titanate  (LaTiO$_3$)  is one of the
most puzzling materials in the family of transition metals (TM) of
the perovskite structure (see, e.g., Ref. \onlinecite{Khom02}).
Each of the Ti ions in this material  has a single valent electron
in one of the three-fold degenerate $t_{2g}$ orbitals, causing the
ground state to be enormously degenerate. Whereas in the majority
of the TM perovskites the conventional Jahn-Teller distortion
lifts the orbital degeneracy, such distortions are believed to be
small in the lanthanum titanate, \cite{Keim} and will be assumed
to be zero for the purposes of the present paper. Here we pose the
following question: is there an alternative mechanism to the
Jahn-Teller one, which is capable of removing the orbital
degeneracy?

This question has already come up in materials with
doubly-degenerate $e_{2g}$ valent electrons. In these materials,
there exists an {\it orbitally ordered phase}, with coherent
excitations (`orbitons'). \cite{Khom02,KKO} These excitations were
recently observed  in LaMnO$_3$. \cite{Sait}  The notion of an
orbital disordered, {\it liquid},  state was developed in Ref.
\onlinecite{IYN} for the hole doped LaMnO$_3$ with a high hole
concentration, where the antiferromagnetic (AF) insulating state
gives way to a ferromagnetic metallic phase. Theoretically, the
orbitons in doped materials are constructed similarly to spinons,
which appear within the framework of the $t-J$ model for the
hole-doped cuprates. \cite{Band} Another scenario of quantum
liquid formation was proposed for the insulating KCuF$_3$.
\cite{FOZ} According to Ref. \onlinecite{FOZ}, the orbital
fluctuations (random changes of the bond directions) may be the
source for the ``quantum melting" of the magnetic order, because
of the strong interaction between orbital and spin degrees of
freedom. Melting, in this scenario, means the disappearance of a
long-range AF order in favor of a {\it spin liquid} state.

The lifting of the  degeneracy of the  $t_{2g}$ orbital triplet in
LaTiO$_3$ is a more difficult problem. In this case, according to
the analysis of Kugel and Khomskii (KK), \cite{KK75} the spin and
the orbital degrees of freedom are not separated. Hence, the spin
excitations are necessarily intimately involved in the formation
of the orbital excitations and {\it vice versa}. A possible
scenario of orbital liquid formation in LaTiO$_3$ has been
proposed in Ref. \onlinecite{KM}. These authors have conjectured
the appearance of a resonance between the spin and the orbital
excitations. To describe this resonance, they have used mean-field
decoupling, generalizing the approach proposed in Ref.
\onlinecite{Band} for the spin liquid state in the two-dimensional
$t-J$ model. This decoupling leads inevitably to the  conclusion
that the resonance states should give a contribution linear in the
temperature to the low-energy specific heat. Such a
Sommerfeld-like contribution signals the lifting of the orbital
degeneracy by forming a quantum liquid. Unfortunately, a direct
experimental check of this prediction in LaTiO$_3$ has yielded
negative results. \cite{Frit02}

In the present paper, the main idea of Khaliullin and Maekawa
\cite{KM} is rehabilitated by means of a more refined approach,
which goes beyond the mean-field decoupling. We use the methods
recently tested in the theory of Kondo lattices, where the spin
liquid state is described as short-range correlations of the RVB
type, which arise due to a non-local exchange through the Kondo
screening clouds between the magnetic ions. \cite{KK02} We show
that the {\it two-magnon} excitations play a similar role in the
formation of the orbital liquid correlations.

Our analysis is based on the KK Hamiltonian \cite{KK75} (see
below). In its simplest symmetric form, this Hamiltonian {\it does
not} support long-range magnetic order at any non-zero
temperature, due to some hidden symmetries.\cite{ABH}   However,
the Hamiltonian of the real LaTiO$_3$ contains many additional
(possibly small) terms, which restore long range AF spin order
below some N\'eel temperature $T_N$, as indeed observed
experimentally.\cite{Keim}
 In the
following, we therefore {\it assume} the existence of this ordered
magnetic state, and concentrate on its effects on the low
temperature dynamics of the orbital degrees of freedom. Once the
magnetic order is stabilized, we expect these dynamics at
temperatures $T \ll T_N$ to be described by the KK Hamiltonian, as
discussed below.   Indeed, we find heavily damped orbital
excitations, which form a continuum whose energy scale is the same
as that of the spin excitations, such that both orbital and spin
excitations merge into a common continuum. As a result, a direct
experimental observation of the orbital dynamics alone, at low
temperatures, is hardly possible.

The paper is organized as follows. In the next section we review
the derivation of the KK Hamiltonian, and set the basis for our
perturbation expansion. The latter is carried out in Sec. III,
where the orbital spectrum is obtained and discussed. Section IV
includes our conclusions.

\section{The Kugel-Khomskii  Hamiltonian}

Following the KK approach, \cite{KK75} we consider a simplified
version of the titanate, concentrating only on the sublattice of
Ti ions. Each Ti ion possesses a single electron in one of the
three-fold degenerate $t_{2g}$ orbitals of the $d$-shell.
Including only a single Coulomb-Hubbard  on-site repulsion on the
Ti ion, with energy $U$, the simple initial Hamiltonian reads
\begin{eqnarray}
{\cal H}&=&\frac{U}{2}\sum_{i}\sum_{mm'}\sum_{\sigma\sigma '}
n_{im\sigma}n_{im'\sigma '}(1-\delta_{mm'}\delta_{\sigma\sigma
'})\nonumber\\
&+&\sum_{\langle ij\rangle}\sum_{mm'}\sum_{\sigma}t_{ij}^{mm'}
d^{\dagger}_{im\sigma}d_{jm'\sigma},\ \ n_{im\sigma}\equiv
d^{\dagger}_{im\sigma}d_{im\sigma}, \label{mod}
\end{eqnarray}
where  $\langle ij\rangle$ denotes nearest-neighbor pairs. Here,
$d^{\dagger}_{im\sigma}$ creates an electron of spin $\sigma$ at
site $i$, in one of the $t_{2g}$ levels, denoted $m$, and
$n_{im\sigma}$ is the number operator in the $m$ orbital on site
$i$ with  spin index $\sigma$.  Kugel and Khomskii \cite{KK75}
treated this Hamiltonian in second-order perturbation theory,
assuming that the overlap integrals $t_{ij}^{mm'}$ are much
smaller than $U$, in order to obtain an effective Hamiltonian
describing the spin interactions among the Ti ions. Operating
within strictly cubic symmetry, KK assumed that
\begin{eqnarray}
t_{ij}^{mm'}=\delta_{mm'}t_{ij}^{m},
\end{eqnarray}
where $t_{ij}^{m}\equiv t$ and differs from zero only when
\begin{eqnarray}
\langle ij\rangle \ \mbox{is along the $z$ axis and}\ m&=&d_{yz}\
{\rm or}\
d_{zx},\nonumber\\
\langle ij\rangle \ \mbox{is along the $y$ axis and}\ m&=&d_{xy}\
{\rm or}\
d_{yz},\nonumber\\
\langle ij\rangle \ \mbox{is along the $x$ axis and}\ m&=&d_{zx}\
{\rm or}\ d_{xy}.\label{d}
\end{eqnarray}
The resulting exchange Hamiltonian, for a bond along the
$z$-direction, reads
\begin{eqnarray}
{\cal H}^{z}_{\rm KK}&=&- \frac{J}{2} \sum_{i} (a^\dagger_{i}a_{i}
+b^{\dagger}_{i}b_{i})\nonumber\\
& +&J\sum_{\langle ij\rangle_{z}}\bigl (\frac{1}{4}+{\bf
S}_{i}\cdot{\bf S}_{j}\bigr )Y^{z}_{\langle ij\rangle},\label{Hz}
\end{eqnarray}
where ${\bf S}_{i}$ is the spin operator on site $i$, and
\begin{eqnarray}
J=2t^{2}/U
\end{eqnarray}
sets the energy scale of the spin excitations. The orbital degrees
of freedom are now described by the operators $a^{\dagger}_{i}$
($b^{\dagger}_{i}$, $c^{\dagger}_{i}$), which create a spinless
fermion (termed `orbiton') in the $yz$ ($zx$, $xy$) orbital on
site $i$. The intermingling of the orbiton degrees of freedom with
those of the spin is contained in the second term of Eq.
(\ref{Hz}), in which KK introduced the orbiton operator product
$Y^{z}$, given by
\begin{eqnarray}
 Y^{z}_{\langle ij\rangle}=
a^{\dagger}_{i}a_{i}a^{\dagger}_{j}a_{j}
+b^{\dagger}_{i}b_{i}b^{\dagger}_{j}b_{j}
+a^{\dagger}_{i}b_{i}b^{\dagger}_{j}a_{j}+
b^{\dagger}_{i}a_{i}a^{\dagger}_{j}b_{j}.\label{Y}
\end{eqnarray}
The expressions for ${\cal H}^{x}_{\rm KK}$ (for bonds along the
$x$-direction) and for ${\cal H}^{y}_{\rm KK}$ (for bonds along
the $y$-direction) are derived from Eq. (\ref{Hz}) by cyclic
permutations of the operators $a$, $b$, and $c$. The full exchange
Hamiltonian is the sum of the three,
\begin{eqnarray}
{\cal H}_{\rm KK}={\cal H}^{x}_{\rm KK}+{\cal H}^{y}_{\rm KK} +
{\cal H}^z_{\rm KK}. \label{fullH}
\end{eqnarray}
Then, the first term in Eq. (\ref{Hz}), when added together with
the corresponding  terms in ${\cal H}^y_{\rm KK}$ and ${\cal
H}_{\rm KK}^x$ yields twice the total number of electrons in the
system,
\begin{eqnarray}
 N=\sum_i \Bigl
(a^\dagger_ia_{i} + b^\dagger_i b_i + c^\dagger_i c_i \Bigr
),\label{N}
\end{eqnarray}
 which  is just a constant.

Contrary to the situation in the manganites, the spin and charge
degrees of freedom in the titanate are not separated. Moreover,
the orbiton-spin interaction has a peculiar structure of two
fermions interacting with one magnon (see below). To explore the
outcome of this complicated situation, we proceed as follows.
Firstly, we work well below T$_{N}$. Then the spin operators may
be treated by the linearized Holstein-Primakoff transformation:
assuming a two sublattice bipartite antiferromagnetic order, with
moments along the magnetic $z$-direction, we write
\begin{eqnarray}
X_{\langle ij\rangle}& = &{\bf S}_i \cdot{\bf S}_j + \frac{1}{4}
\nonumber\\
& = &\frac{1}{2}\Bigl (B^\dagger_i B_i + B^\dagger_j B_j + B_i B_j
+ B^\dagger_i B^\dagger_j\Bigr). \label{XX}
\end{eqnarray}
Here, $B^\dagger_i$ is a boson operator which creates a magnon
excitation (related to $S_{ix}+iS_{iy}$). Secondly, the KK
exchange Hamiltonian, together with Eq. (\ref{XX}), will be
treated as a perturbation, acting on the zeroth order Hamiltonian.
The latter consists of the free orbiton Hamiltonian, and the free
spin Hamiltonian. For the first, we impose a constraint ensuring
that each site is occupied by a single charge carrier. This
constraint is applied {\it globally}. For the latter, we introduce
an infinitesimally  small staggered magnetic field in the magnetic
$z$-direction, $h_i = (-1)^i h$. Hence,
\begin{eqnarray}
{\cal H}_0& = &{\cal H}_O + {\cal H}_S,
\end{eqnarray}
with
\begin{eqnarray}
{\cal H}_O& = & - \lambda \sum_i \Bigl (a^\dagger_i a_i +
b^\dagger_i b_i + c^\dagger_i c_i\Bigr),
\label{OH}\\
{\cal H}_S& = & \sum_i h_i \Bigl ( \frac{1}{2} - B^\dagger_i
B_i\Bigr ). \label{SH}
\end{eqnarray}
Note that the chemical potential $\lambda$ includes the constant
term of the original KK Hamiltonian (\ref{fullH}) [{\it cf}.  Eq.
(\ref{N})]. In the next section we use these definitions in order
to construct a thermodynamic perturbation theory in terms of the
Matsubara Green functions.

\section{Perturbation expansion for the orbiton spectrum}

The orbiton spectrum is determined by the self-energy of the
orbiton Green function. Here we calculate the orbiton Green
function, which is diagonal in the orbiton operators, that is,
\begin{eqnarray}
G_{ij}^{\mu\mu}(\tau)= \langle T_\tau
a_{i\mu}(\tau)a_{j\mu}^\dagger(0)\rangle ,
\label{OG}
\end{eqnarray}
where $i$ and $j$ denote lattice sites, $T_{\tau}$ is the
time-ordering operator, and $a_\mu=a,b,$ or $c$. This Green
function satisfies the Dyson equation
\begin{eqnarray}
G_{ij}^{\mu\mu} = g_{0}\left( \delta_{ij} + \sum_{\nu}\sum_{l}
M^{\mu\nu}_{il} G_{lj}^{\nu\nu} \right), \label{dyson}
\end{eqnarray}
where $g_{0}$ is the `bare' orbiton Green function,
\begin{eqnarray}
g_{0}(\epsilon_{m})=(i\epsilon_m + \lambda )^{-1},
\end{eqnarray}
with $\epsilon_m$ being the Matsubara frequency. As we show below,
the self-consistent calculation of the orbiton self-energy,
$M^{\mu\nu}_{il}$, requires knowledge of the magnon Green
function, which may be represented by the matrix
\begin{eqnarray}
\widehat{D}_{ij}(\tau )=\left[ \begin{array}{cc} \langle T_\tau
B_{i}(\tau )B_j^\dagger(0)\rangle
& \langle T_\tau B_{i}(\tau )B_j(0)\rangle  \\
\langle T_\tau B_i(\tau)^\dagger B_j^\dagger(0) \rangle & \langle
T_\tau B_i^\dagger(\tau) B_j(0) \rangle
\end{array}
\right ].
\end{eqnarray}
Unlike the standard case of a linear fermion-boson interaction,
the KK Hamiltonian (\ref{fullH}) contains six-tail interaction
vertices, of four types (see Fig. 1). These are denoted $V_1$,
$V_2$, $V_3$, and $V_4$ and are given by
\begin{eqnarray}
V_1 & = & - JB_i^\dagger B_i
a^\dagger_{i\mu}a_{\ell\mu}a^\dagger_{\ell\nu}a_{i\nu},
\nonumber \\
V_3 & = & - J B_i^\dagger B_\ell^\dagger
a^\dagger_{i\mu}a_{\ell\mu} a^\dagger_{\ell\nu}a_{i\nu},\label{V1}
\end{eqnarray}
$V_2$ has the same form as $V_1$ with $i$ replaced by $\ell$, and
$V_4 = V_3^\dagger$. Hence the KK Hamiltonian involves a
fermion-magnon interaction, which is {\it quadratic} in the magnon
operators and {\it quartic} in the fermion operators. This feature
is the basis for our results in the following.

\vspace{1cm}

\begin{figure}
\leavevmode \epsfclipon \epsfxsize=8truecm
\vbox{\epsfbox{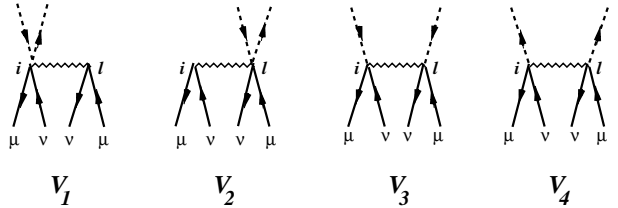}} \vspace{1cm} \caption{The four
magnon-orbiton vertices. The wavy lines denote the interaction,
full (dotted) lines denote the orbiton (magnon)
operators.}\label{fig1}
\end{figure}

\subsection{Lowest-order mean-field approximation }

To first order, the perturbation expansion corresponds to the
mean-field approximation for the self-energy of both the orbitons
and the magnons. Then, these self-energies can be obtained by
decoupling the product of magnon operators and orbiton operators
appearing,  e.g, in Eq. (\ref{Hz}). This procedure is carried out
as follows. Ignoring a constant, one first writes
\begin{eqnarray}
{\cal H}^{z}\equiv \sum_{\langle ij\rangle_{z}}X_{\langle
ij\rangle}Y^{z}_{\langle ij\rangle}\sim\sum_{\langle
ij\rangle_{z}}\Bigl [\langle Y\rangle X_{\langle
ij\rangle}+\langle X\rangle Y^{z}_{\langle ij\rangle}\Bigr
],\label{first}
\end{eqnarray}
where the orbiton operators $Y$ and the magnon operators $X$ are
given in Eqs. (\ref{Y}) and (\ref{XX}), respectively. The notation
$\langle\rangle$ stands for the thermodynamic average of the
relevant operators. A similar decoupling is written down for
${\cal H}^{x}$ and ${\cal H}^{y}$. When these three approximate
Hamiltonians are summed together, the resulting Hamiltonian
separates into two terms, ${\cal H}_{1}$ and ${\cal H}_{2}$, the
first pertaining to the magnons, and the second to the orbitons.

Let us first consider ${\cal H}_{1}$ and the magnon dynamics it
implies. The orbiton operator product contained in $\langle
Y\rangle$ is further decoupled, by introducing the thermal
averages, which will be determined self-consistently (see below),
\begin{eqnarray}
n_\mu& = &\langle a_{i\mu}^\dagger a_{i\mu}\rangle,\ {\rm with}\ n
\equiv n_a = n_b = n_c \equiv 1/3,
\nonumber\\
\Delta_\mu& = &\langle a^\dagger_{i\mu} a_{j\mu}\rangle ,\ {\rm
with}\ \Delta \equiv \Delta_a = \Delta_b = \Delta_c,\  i \ne j.
\label{ndelta}
\end{eqnarray}
In other words, we assume an equal average occupation of all three
orbitals (i.e., no orbital long range order), and consider only
the possible ``ordering" of the off-diagonal orbital parameter,
$\Delta$, which represents the delocalization of the orbital
degrees of freedom. Then ${\cal H}_{1}$ takes the form
\begin{eqnarray}
{\cal H}_{1} = \frac{\tilde{J}}{2} \sum_{\langle ij \rangle}\Bigl
( B^\dagger_i B_i + B^\dagger_j B_j + B_i B_j + B^\dagger_i
B^\dagger_j \Bigr), \label{magnon}
\end{eqnarray}
with the coupling constant
\begin{equation}
\tilde{J} = 2 J (n^2 - 2 \Delta^2). \label{JTILD}
\end{equation}
Thus, ${\cal H}_1$ reduces to the usual spin-wave
antiferromagnetic Hamiltonian, with a renormalized exchange
coefficient. As usual, this Hamiltonian is diagonalized in Fourier
space, utilizing the Holstein-Primakoff rotation:
\begin{eqnarray}
B_{\bf q} =C_{\bf q}\beta_{\bf q} + S_{\bf q} \beta^\dagger_{-{\bf
q}},
\end{eqnarray}
where
\begin{eqnarray}
C_{\bf q} \equiv \cosh W_{\bf q},\ \ S_{\bf q} \equiv \sinh W_{\bf
q},\label{CS}
\end{eqnarray}
such that
\begin{eqnarray}
- \tanh 2W_{\bf q} = \frac{1}{3}(\cos q_x + \cos q_y + \cos
q_z)\equiv \Phi_{\bf q}.
\end{eqnarray}
Hence, the magnon Green function takes the form
\begin{eqnarray}
{\cal D}_{\bf q}(\omega_n)=(i\omega_n - \omega_{\bf q})^{-1},
\end{eqnarray}
with the magnon frequency
\begin{eqnarray}
\omega_{\bf q} = 3 \tilde{J} \sqrt{1 - \Phi_{\bf q}^{2}}.
\label{omega}
\end{eqnarray}

\begin{figure}
\leavevmode \epsfclipon \epsfxsize=8.3truecm
\vbox{\epsfbox{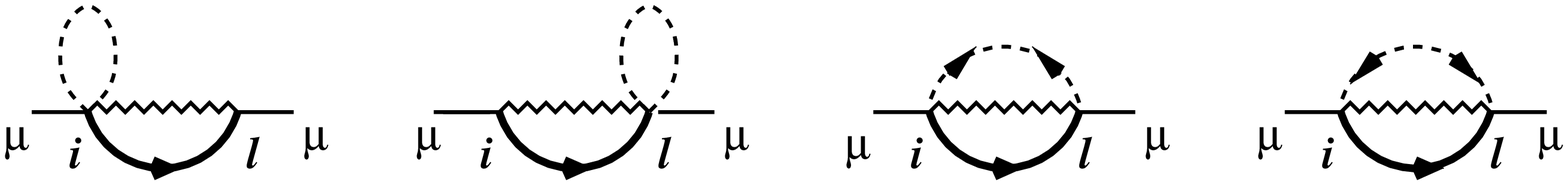}} \vspace{1cm} \caption{The first-order
diagrams contributing to the self-energy of the orbiton Green
function.       }\label{fig2}
\end{figure}

Within the mean field approximation of Eq. (\ref{first}), the
energy scale of the orbiton Hamiltonian ${\cal H}_2$ is set by
$\langle X\rangle$, which can be now obtained using the magnon
dispersion (\ref{omega}),
\begin{equation}
\langle X \rangle =\frac{3}{2N}\sum_{\bf q}\left [- 1 + (2N_{\bf
q} + 1) \sqrt{1 - \Phi_{\bf q}^2}~ \right ], \label{X}
\end{equation}
with the magnon occupation numbers $N_{\bf q} =\bigl (\exp
(\beta\omega_{\bf q}) - 1\bigr )^{-1}$. Since the operator
$Y^z_{\langle ij \rangle}$ still involves products of four
operators, one needs to apply further contractions. It is then
convenient to calculate the orbiton self-energy using the diagrams
shown in  Fig. \ref{fig2}. From these diagrams one finds the first
order self-energy
\begin{eqnarray}
M_{1,\mu}({\bf k}) = 4 J \langle X \rangle [n - \Phi^\mu_{\bf k}
\Delta ] , \label{self}
\end{eqnarray}
with the form factors
\begin{eqnarray}
\Phi^{a}_{\bf k}=\cos k_{y}+\cos k_{z},
\end{eqnarray}
and analogous expressions for $\Phi^b$ ($y,z\rightarrow z,x$) and
$\Phi^c$ ($y,z\rightarrow x,y$). It therefore follows that the
orbiton Green function, in this approximation, becomes
\begin{eqnarray}
G^{\mu\mu}_{\bf k}(\epsilon)=\frac{1}{\epsilon -E^\mu_{\bf k}},
\label{G1}
\end{eqnarray}
where
\begin{eqnarray}
E^\mu_{\bf k} = 4 J \langle X\rangle [n - \Phi^\mu_{\bf k} \Delta]
- \lambda , \ \mu =a,b,c. \label{E}
\end{eqnarray}

\vspace{0.5cm}
\vspace{0.5cm}

It is thus seen that a finite value of $\Delta $ implies a finite
dispersion of the orbiton modes, with an orbiton band width $16J
\langle X \rangle \Delta$. Moreover, this band-width is
temperature-dependent, due to the temperature dependence of the
magnon occupations, contained in $\langle X\rangle$, Eq.
(\ref{X}), and due to the temperature dependence of $\Delta$
itself. For a given wave-vector ${\bf k}$, the three energies
$\{E^\mu_{\bf k}\}$ are generally different, so that the
degeneracy of the $t_{2g}$ band is removed. Indeed, were we to
terminate the calculation at this level, keeping only first-order
contributions, then the orbiton averages, Eqs. (\ref{ndelta}),
would have to be found self-consistently using the Green function
(\ref{G1}). It is very illuminating to examine those
self-consistency requirements. Equation (\ref{G1}) yields
\begin{eqnarray}
f^\mu_{\bf k} \equiv \langle a^\dagger_{\mu{\bf k}}a_{\mu{\bf k}}
\rangle = f(E^\mu_{\bf k}) \equiv \frac{1}{e^{\beta E^\mu_{\bf
k}}+1},\ \mu =a,b,c. \label{f}
\end{eqnarray}
Then the self-consistency equations read
\begin{eqnarray}
1 & = & \frac{1}{N} \sum_\mu \sum_{\bf k} f^\mu_{\bf k}, \ \ \
\Delta (T)=\frac{1}{6N}\sum_{\mu}\sum_{\bf k} \Phi^{\mu}_{\bf k}
f^{\mu}_{\bf k}, \label{self1}
\end{eqnarray}
where the first equation fixes the chemical potential
$\lambda(T)$, so that the three-fold-degenerate orbiton band is
1/3 filled, and the second determines $\Delta$.

Evidently, Eqs. (\ref{self1}) are satisfied by setting $\Delta
=0$, (since then the Fermi function is independent of {\bf k}, and
$\sum_{\bf k}\Phi^\mu_{\bf k} $ vanishes). Let us now examine at
which temperature $\Delta $ begins to deviate from zero. To this
end we expand Eqs. (\ref{self1}) in powers of $\Delta$. The first
self-consistency equation yields (at $\Delta=0$)
\begin{eqnarray}
4 J \langle X\rangle (T) n - \lambda(T) = \frac{4}{3} J \langle X
\rangle (T) - \lambda (T) = \frac{1}{\beta}\ln 2,
\end{eqnarray}
while the second allows a non-zero $\Delta$ only for temperatures
lower than the solution of
\begin{eqnarray}
\frac{1}{\beta}=\frac{4}{9}J \langle X \rangle (T).
\end{eqnarray}
Using here the explicit expression for $\langle X \rangle$, Eq.
(\ref{X}), and  the results Eqs. (\ref{JTILD}) and (\ref{omega})
for the magnon frequency, we find
\begin{eqnarray}
\frac{3}{2\beta J} = -1 + \frac{3}{2\beta J} \frac{1}{N} \sum_{\bf
q} \beta\omega_{\bf q}~\coth\frac{\beta\omega_{\bf q}}{2}.
\end{eqnarray}
This is an implicit equation for the temperature at which $\Delta
$ begins to deviate from zero. Examination of this result reveals
that this occurs  at $k_BT$ of order $J$. A similar order of
magnitude is found from the spin-wave approximation for the N\'eel
temperature (above which the staggered magnetization
$(\frac{1}{2}-B_i^\dagger B_i)$ vanishes). Both of these estimated
temperatures are probably above the real N\'eel temperature, which
is determined by small symmetry breaking terms which are not
contained in the simple KK model. Hence, in the whole range below
the real $T_N$  we expect to have a finite $\Delta (T)$, and we do
not expect any phase transition at which $\Delta$ vanishes while
the spins are still ordered.  Even if this statement is not valid
for all $T<T_N$, we still expect it to hold for sufficiently low
temperatures. In the next section we examine the dynamical
corrections to the orbiton spectrum, brought about by the
second-order corrections, and examine whether and how the above
conclusion is modified.

\subsection{Beyond the mean-field approximation}

\begin{figure}
\leavevmode \epsfclipon \epsfxsize=8truecm
\vbox{\epsfbox{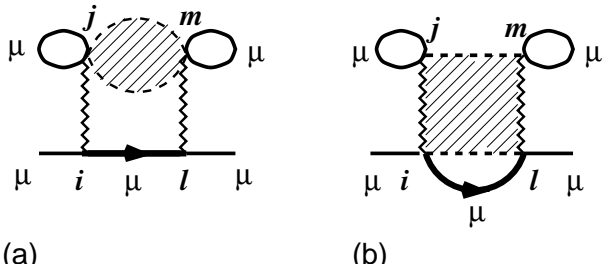}} \vspace{1cm} \caption{Second-order
diagrams contributing to the self-energy of the orbiton Green
function. }\label{fig3}
\end{figure}

The next order contribution to the orbiton self-energy,
$M_{2,\mu}({\bf k},\epsilon )$, requires expansion to second order
in the KK Hamiltonian, i.e. to second order in the diagrams listed
in Fig. 1. Instead of going through the details of all the
diagrams which contribute to $M_{2,\mu}$, we give here the details
of only two diagrams,  which describe orbiton propagation assisted
by two-magnon processes (Fig. \ref{fig3}). The first of these two
diagrams is generated by the vertices $V_1$ and $V_2$ in Fig.
\ref{fig1}, whereas the second is generated by the vertices $V_3$
and $V_4$. Both diagrams contain a single propagating fermion
line. Each of the other two fermion tails (of each vertex) is
rolled into a bubble, to give the factor $n^2$. Other diagrams
involve three internal fermion lines, but do not change the
qualitative features discussed below.

\vspace{1cm}

A lengthy calculation of the contribution of the two diagrams
displayed in Fig. \ref{fig3} yields (after performing the
analytical continuation for the Matsubara frequencies)
\begin{eqnarray}
&&M^{(1)}_{2,\mu} ({\bf k},\epsilon ) = \left
(\frac{nJ}{N}\right)^2 \sum_{\bf pq}(\Phi_{\bf p}^{\mu})^2
\nonumber\\
&&\times\left [C^+_{{\bf q},{\bf p} + {\bf q}} \frac{N_{\bf q}(1 +
N_{{\bf p} + {\bf q}}) + f^\mu_{{\bf k} - {\bf p}} (N_{{\bf p} +
{\bf q}} - N_{\bf q})}{\epsilon - E^\mu_{{\bf k} - {\bf p}} +
\omega_{\bf q} - \omega_{{\bf p} + {\bf q}}}\right.
\nonumber\\
& + & \frac{1}{2}C^-_{{\bf q},{\bf p} + {\bf q}} \left(
\frac{N_{\bf q}N_{{\bf p} + {\bf q}} + f^{\mu}_{{\bf k} - {\bf
p}}( 1 + N_{{\bf p} + {\bf q}} + N_{\bf q})}{ \epsilon -
E^\mu_{{\bf k} - {\bf p}} + \omega_{\bf q} + \omega_{{\bf p} +
{\bf q}}}\right.
\nonumber\\
&+& \left. \left.\frac{(1 + N_{\bf q})(1 + N_{{\bf p} + {\bf q}})
- f^\mu_{{\bf k} - {\bf p}}(1 + N_{{\bf p} + {\bf q}} + N_{\bf
q})}{\epsilon - E^\mu_{{\bf k} - {\bf p}} - \omega_{\bf q} -
\omega_{{\bf p} + {\bf q}}} \right )\right ],\nonumber\\
\label{2a}
\end{eqnarray}
for the first diagram, and
\begin{eqnarray}
&&M^{(2)}_{2,\mu}({\bf k},\epsilon ) = \left( \frac{nJ}{N}\right
)^2 \sum_{\bf pq}\left(\frac{\Phi^\mu_{\bf q} + \Phi^\mu_{{\bf
p} + {\bf q}}}{2}\right )^{2}\nonumber\\
&& \times \left [C^-_{{\bf q},{\bf p} + {\bf q}} \frac{N_{\bf q}(1
+ N_{{\bf p} + {\bf q}}) + f^\mu_{{\bf k} - {\bf p}}(N_{{\bf p} +
{\bf q}} - N_{\bf q})}{\epsilon - E^\mu_{{\bf k} - {\bf p}} +
\omega_{\bf q} - \omega_{{\bf p} + {\bf q}}}\right.
\nonumber\\
& + &\frac{1}{2} C^+_{{\bf q},{\bf p} + {\bf q}} \left(
\frac{N_{\bf q} N_{{\bf p} + {\bf q}} + f^\mu_{{\bf k} - {\bf
p}}(1 + N_{{\bf p} + {\bf q}} + N_{\bf q})}{\epsilon -
E^{\mu}_{{\bf k} - {\bf p}} + \omega_{\bf q} + \omega_{{\bf p} +
{\bf q}}}\right.
\nonumber\\
& + & \left.\left.\frac{(1 + N_{\bf q})(1 + N_{{\bf p} + {\bf q}})
- f^\mu_{{\bf k} - {\bf p}}(1 + N_{{\bf p} + {\bf q}} + N_{\bf
q})}{\epsilon - E^\mu_{{\bf k} - {\bf p}} - \omega_{\bf q} -
\omega_{{\bf p} + {\bf q}}} \right )\right ],
\nonumber\\
\label{2b}
\end{eqnarray}
for the second diagram. Here we have introduced the magnon
coherence factors,
\begin{eqnarray}
C^+_{{\bf q},{\bf p} + {\bf q}}& = &\left (C_{\bf q} C_{{\bf p} +
{\bf q}} + S_{\bf q} S_{{\bf p} + {\bf q}} \right )^2,
\nonumber\\
C^-_{{\bf q},{\bf p} + {\bf q}}& = &\left( S_{\bf q} C_{{\bf p} +
{\bf q}} + C_{\bf q} S_{{\bf p} + {\bf q}}\right)^2,
\end{eqnarray}
which, upon using Eq. (\ref{CS}), take the simpler form
\begin{eqnarray}
C^\pm_{{\bf q},{\bf p} + {\bf q}} & = &\left \{
\begin{array}{c}
\cosh^2 (W_{\bf q} + W_{{\bf p} + {\bf q}})\\{\sinh}^2 (W_{\bf q}
+ W_{{\bf p} + {\bf q}})
\end{array}\right. .
\end{eqnarray}

With the second-order corrections to the orbiton self-energy, the
orbiton Green function becomes
\begin{eqnarray}\label{G2}
G^{\mu\mu}_{\bf k}(\epsilon) = \frac{1}{\epsilon - E^\mu_{\bf k} -
M_{2,\mu}({\bf k}, \epsilon )},
\end{eqnarray}
where $M_{2,\mu}({\bf k},\epsilon )=M_{2,\mu}^{(1)}({\bf
k},\epsilon ) + M_{2,\mu}^{(2)}({\bf k}, \epsilon )+...$. We now
concentrate on the contributions of Eqs. (\ref{2a}) and (\ref{2b})
to this Green function. Clearly, these contributions yield a
finite, temperature-dependent imaginary part to the orbiton
dispersion, namely, the orbiton modes acquire a
temperature-dependent damping, as expected. The new dispersion law
is obtained from the poles of the Green function (\ref{G2}),
\begin{eqnarray}
z - E^\mu_{\bf k} = M_{2,\mu}({\bf k},z ),
\end{eqnarray}
where $z$ is a complex variable. Examining Eqs. (\ref{2a}) and
(\ref{2b}) one may draw the following conclusions. \noindent (i)
The real part of the second-order self-energy, $R({\bf k},\epsilon
) = {\rm Re}~ M_{2,\mu}({\bf k},\epsilon )$, as  function of
$\epsilon$, contains three components which come from the
two-magnon processes. The lowest one arises from the two-magnon
emission processes and is dominant in the interval $- 2 \omega_D
\leq \epsilon\leq 0$, where $\omega_D$ is the maximal magnon
frequency [see Eq. (\ref{omega})]. The second results from the
processes in which one magnon is emitted and one is absorbed,
which occur in the interval $- \omega_D \leq \epsilon \leq
\omega_D$. The third component comes from the two-magnon
absorption processes and prevails at positive frequencies,
$0\leq\epsilon\leq 2\omega_{D}$. It follows that $R({\bf
k},\epsilon )$ has a significant magnitude in the entire range
$-2\omega_D < \epsilon < 2\omega_D$, and falls down rapidly
outside this interval. The magnitude of different contributions to
$R({\bf k},\epsilon )$ is regulated by the magnon occupation
numbers. In particular, the mixed emission-absorption processes
are frozen out at $T\to 0$. One may conclude  that the
superposition of the three types of the two-magnon processes
results in a complicated behavior of $R({\bf k},\epsilon)$. This
quantity may have several zeros and peaks within the above energy
interval; the heights of these peaks can be estimated as $\sim
\kappa J$, with a numerical coefficient $\kappa <1$. \noindent
(ii) The imaginary part of the orbiton self-energy, which comes
from the second order terms, $\Gamma({\bf k},\epsilon ) = {\rm
Im}~ M_{2,\mu}({\bf k},\epsilon )$, behaves similarly to the real
part; expressions (\ref{2a}) and (\ref{2b}) do not show that
$\Gamma$ is essentially smaller than $R$ in the whole energy
interval. \noindent (iii) As a result, the orbitons become
ill-defined, heavily damped, quasi-particles. Their off-diagonal
orbiton parameter $\Delta$, and the related orbiton band-width,
should be again determined self-consistently, using the refined
orbiton Green function Eq. (\ref{G2}),
\begin{eqnarray}
\Delta (T)=-\frac{1}{6N}\sum_{\bf
k}\sum_{\mu}\int_{-\infty}^{\infty}
\frac{d\epsilon}{2\pi}\Phi^{\mu}_{\bf k}{\rm Im} G^{\mu\mu}_{\bf
k}(\epsilon)f(\epsilon ),\label{SELF2}
\end{eqnarray}
in place of the second of Eqs. (\ref{self1}).

Without going into the details of the solutions of the complicated
self-consistent equation for $\Delta (T)$, Eq. (\ref{SELF2}), it
is clear that going beyond the simple mean-field calculation
changes drastically the nature of the orbiton excitations. Whereas
in the first order we have obtained a coherent orbiton spectrum
with a well-defined wave vector (appearing due to the magnetic
order), this picture changes drastically with the second-order
corrections. The interaction with the spin wave excitations
extends the effective width of the orbiton band and makes it
asymmetric relative to its center of gravity. Moreover, the
orbiton excitations lose their coherence by acquiring a
considerable life-time (given by the imaginary part of the
self-energy). Those excitations are thus transformed into a
structureless ``liquid-like" continuum. Clearly, more complicated
processes involving  higher-orders of the orbiton-magnon
interaction, are not expected to modify this picture.

\section{Concluding remarks}

We have proposed in this paper a mechanism capable of  lifting the
degeneracy of the  $t_{2g}$ orbitals  in the titanate. In our
scenario, the orbital degeneracy is removed due to the interplay
between the (ordered) spin and the (disordered) orbital degrees of
freedom. We have obtained three dispersive orbiton bands, with a
temperature-dependent band-width and a structureless spectrum,
that exist together with the coherent spin fluctuations at low
temperatures.  We have found that the second-order perturbation
corrections transform the coherent orbiton spectrum, $E_{\bf
k}^\mu$, [see Eq. (\ref{E})] obtained within the simplest
mean-field theory, into an incoherent continuum of heavily damped
excitations, which may be termed an `orbital liquid'. This, in
turn, may explain the experimental observation, \cite{Frit02} that
the low-temperature heat capacity arises from the spin wave
excitations alone.

The incoherent continuum state of the orbital excitations is
dominated at low temperatures by the two-magnon emission peak in
the orbiton self-energy. The effect of this peak is to shift the
weight of the orbiton spectrum into negative energies, thus making
the orbiton liquid state thermodynamically preferable as compared
to the fully-degenerate orbiton state that prevails when
$\Delta=0$, or in the absence of the coupling with the spin
degrees of freedom. In particular, our process of degeneracy
lifting may possibly overcome the Jahn-Teller mechanism.
\cite{imada} In the latter, the energy gain is restricted by the
energy of the Jahn-Teller coupling strength, which is definitely
smaller than the exchange coupling, which determines the energy
gain in our case.

Finally we mention that, within the framework of the KK
Hamiltonian in cubic symmetry, the orbiton-magnon interaction
cannot lead to a magnetic anisotropy, that is, it cannot generate
a gap in the spin-wave spectrum. To obtain such anisotropies, one
should extend the KK Hamiltonian to include the hybridization of
the oxygen and titanium orbitals, the possible orthorhombic
tilting of the oxygen octahedra, as well as spin-orbit
interactions. Apparently, one will also have to treat the Coulomb
interaction in a more refined way, taking into account
Coulomb-exchange effects.\cite{IYN}

\acknowledgements We acknowledge partial support by the
German-Israeli Foundation (GIF), and the US-Israel Binational
Science Foundation (BSF). Valuable discussions with D. Khomskii
are highly appreciated.


\end{multicols}
\end{document}